\documentstyle[twocolumn,epsf]{jpsj}

\input defphys.inc

\def\runtitle{
Effects of Electron Correlation, Orbital Degeneracy and
Jahn-Teller Coupling in Perovskite Manganites
}
\def\runauthor{Yukitoshi {\sc Motome} and Masatoshi {\sc Imada}}

\def\sf{\rm}

\title{
Effects of Electron Correlation, Orbital Degeneracy and\\
Jahn-Teller Coupling in Perovskite Manganites
}

\author{Yukitoshi {\sc Motome}$^{1}$ and Masatoshi {\sc Imada}$^{2}$
}
\inst{
$^{1}$Department of Physics, Tokyo Institute of Technology,
Oh-okayama 2-12-1, Meguro-ku, Tokyo 152-8551 \\
$^{2}$Institute for Solid State Physics, University of Tokyo,
Roppongi 7-22-1, Minato-ku, Tokyo 106-8666
}
\recdate{August 6, 1998}

\abst{
Roles of Coulomb interaction, orbital degeneracy and Jahn-Teller coupling
in double-exchange models are examined for Mn perovskite oxides.
We study the undoped Mott insulator
as well as metal-insulator transitions by hole doping, and especially
strong incoherence of ferromagnetic metal.
We derive models where
all the spins are fully polarized in two-dimensional planes
as in the experimental indications,
and investigate their ground-state properties
by quantum Monte Carlo method.
At half filling where the number of $e_{g}$ electron
is one per site on average,
the Coulomb interaction opens
a Mott gap and induces a staggered orbital ordering.
The opening of the Mott gap is, however,
substantially slower than the mean-field results
if the Jahn-Teller coupling is absent.
The synergy between the strong correlation and the Jahn-Teller coupling
largely enhances the Mott gap amplitude and reproduces
realistic amplitudes and stabilization energy
of the Jahn-Teller distortion.
Upon doping, the orbital ordering stabilized by the Coulomb interaction
is destroyed immediately.
Toward the metal-insulator transition,
the short-ranged orbital correlation is critically enhanced in metals,
which should be related to strong incoherence of charge dynamics
observed in experiments.
Our model, moreover, exhibits a uniform ordering of
$d_{x^{2}-y^{2}}$ orbital in a wide region of doping
in agreement with experimental indications.
}

\kword{
perovskite manganites, double exchange model, Coulomb interaction,
orbital degeneracy, Jahn-Teller coupling,
Mott gap, orbital ordering, Jahn-Teller distortion,
metal-insulator transition, ferromagnetic metal,
incoherent charge dymanics, quantum Monte Carlo method
}

\begin{document}
\sloppy
\maketitle

`Simple' double-exchange (DE) models, that is, 
models with a single non-interacting conduction band
%for the $e_{g}$ orbital
ferromagnetically coupled with localized spins,
%in the $t_{2g}$ orbital,
have been intensively studied
\cite{Zener1951, Anderson1955, deGennes1960, Kubo1972, Furukawa1994}
to understand physical properties of Mn perovskite oxides,
especially collosal magnetoresistance.
\cite{Jonker1950, Wollan1955, Searle1969, Tokura1994, Ramirez1997}
These models have successfully explained
several experimental aspects at least qualitatively;
for instance,
ferromagnetic metals in a hole-doped region
with a transition to a paramagnetic state by raising temperature, and
a large negative magnetoresistance near the transition.
All these properties are due to 
the so-called double-exchange mechanism;
doped holes tend to gain kinetic energy by aligning localized spins in parallel.
\cite{Zener1951, Anderson1955, deGennes1960}

However, many open problems still remain.
In particular, we focus here the following:
(i) All the materials are the Mott insulator with a charge gap of order of eV
at half filling (one $e_{g}$ electron per Mn site on average),
where two-dimensional (2D) anisotropy realizes in spin and orbital orderings,
and a cooperative Jahn-Teller (JT) distortion.
\cite{Wollan1955, Goodenough1955, Elemans1971, Bocquet1992, Arima1993, Chainani1993}
(ii) Upon doping of holes, a metal-insulator (MI) transition occurs
from the anisotropic Mott insulator to an isotropic ferromagnetic metal
through disordered ferromagnetic insulator.
\cite{Jonker1950, Tokura1994, Urushibara1995}
(iii) In the ferromagnetic metallic state,
charge dynamics shows strong incoherence with tiny Drude weight
even at low temperatures while the linear specific-heat coefficient
$\gamma$ remains small.
\cite{Okimoto1995, Okimoto1997}
All these problems suggest importance of elements neglected
in the simplified models;
Coulomb interaction, orbital degrees of freedom, JT coupling
and so on.

In several recent studies, effects of these factors
previously neglected were also explored;
mean-field approximation,
\cite{Kugel1973, Mizokawa1995, Mizokawa1996, Shiina1997, Maezono1998}
exact diagonalization of small clusters,
\cite{Koshibae1997}
Gutzwiller technique,
\cite{Zang1996}
slave-fermion theory,
\cite{Ishihara1997}
dynamical mean-field theory
\cite{Millis1996, Brito1998}
and perturbation theory.
\cite{TakahashiPREPRINT}
Nevertheless it is still controversial
which of the electron correlation, orbital degeneracy or
the JT coupling is the driving mechanism of experimental properties,
particularly strongly incoherent charge dynamics,
and what are realistic parameters of theoretical models.
These need to fully consider quantum fluctuations.

In this work, we investigate DE models including
Coulomb interaction, orbital degeneracy of conduction bands
and JT coupling.
Especially, we discuss quantum fluctuation
effects of on-site Coulomb interaction
on half-filled states as well as
on MI transitions by hole doping in the ground state.
%In real systems, the Mott gap is order of eV at half filling and
%the Jahn-Teller distortion and orbital ordering occur
%at temperatures around or below 0.1 eV.
%\cite{Bocquet1992, Arima1993, Chainani1993}
%This suggests importance of the Coulomb interaction
%as a principal factor especially in a metal-insulator transition
%because it is essentially a Mott transition.
%\cite{Mott1990, Imada1998}
In order to obtain unbiased results
under strong quantum fluctuation,
we use here quantum Monte Carlo (QMC) technique.
\cite{White1989, Imada1989}
Our results at half filling show that
a strong Coulomb interaction is required to 
understand undoped manganites quantitatively.
Experimentally-observed JT distortions
\cite{Wollan1955, Goodenough1955, Elemans1971}
are reproduced
as a consequence of the synergy between the strong correlation and
the JT coupling.
Nature of the MI transition is investigated in detail;
we show that this transition is a continuous one with
a critical enhancement of orbital correlations,
which may be relevant to incoherent charge dynamics.
\cite{Okimoto1995, Okimoto1997}
A uniform orbital ordering observed in doped materials
\cite{Kawano1998,Akimoto1998}
is also reproduced in our model.

The model we study here consists of three terms as
\begin{equation}
\label{H tot}
{\cal H} = {\cal H}_{\rm el} + {\cal H}_{\rm el-ph} + {\cal H}_{\rm ph}.
\end{equation}
The first term is derived 
in a particular limit of DE models with on-site Coulomb interaction
as well as with twofold degeneracy of $e_{g}$ orbitals.
The original form of ${\cal H}_{\rm el}$ is given by
%\begin{equation}
%\label{H DE2}
$
{\cal H}_{\rm DE} = {\cal H}_{t} + {\cal H}_{\rm int} + {\cal H}_{K},
$
%\end{equation}
where
%\begin{equation}
$
{\cal H}_{t} = \sum_{ij, \nu\nu', \sigma}
t_{ij}^{\nu\nu'} c_{i\nu\sigma}^{\dagger} c_{j\nu'\sigma}
$
%\end{equation}
is the hopping term,
here $c_{i\nu\sigma} (c_{i\nu\sigma}^{\dagger})$ annihilates
(creates) a $\sigma$-spin $e_{g}$ electron
at site $i=1,\cdot\cdot\cdot,N_{\rm S}$ with orbital $\nu=1$ or $2$;
${\cal H}_{\rm int}$ contains on-site interactions
within the doubly-degenerate $e_{g}$ orbitals;
and 
%\begin{equation}
$
{\cal H}_{K} = K \sum_{i} {\bf S}^{e_{g}}_{i}\cdot{\bf S}^{t_{2g}}_{i}
$
%\end{equation}
denotes the Hund's rule coupling between
$e_{g}$ electrons and $t_{2g}$ localized spins.
%Here the indices $i,j = 1,\cdot\cdot\cdot,N_{\rm S}$ and 
%$\nu,\nu' = 1,2$ denote Mn lattice sites and
%doubly-degenerate $e_{g}$ orbitals.
Following previous studies,
\cite{Anderson1955, deGennes1960, Kubo1972, Furukawa1994}
we consider the limit of $K |{\bf S}^{t_{2g}}| \gg t_{ij}^{\nu\nu'}$.
%which is appropriate for many perovskite manganites.
In this limit, ${\bf S}^{e_{g}}_{i}$ aligns parallel to  ${\bf S}^{t_{2g}}_{i}$
in each site, therefore $t_{ij}^{\nu\nu'}$
is renormalized by a relative angle of $t_{2g}$ spins in sites $i$ and $j$.
Moreover, we assume perfect polarization of spins
in the ground state based on the following arguments.
At finite hole doping, it is known that
${\cal H}_{\rm DE}$ shows a metallic ground state
with perfect polarization of spins,
which corresponds well to many doped manganites.
\cite{Urushibara1995, Okimoto1995, Okimoto1997, Shiba1997, 
TakahashiPREPRINT}
At half filling, these materials commonly become
an $A$-type antiferromagnetic insulator
which consists of ferromagnetic layers stacking antiferromagnetically.
\cite{Wollan1955, Goodenough1955, Elemans1971}
These show that the 2D structure of
perfectly-polarized ferromagnetic planes may persist
in both sides of MI transition caused by hole doping.
Therefore, within a 2D plane at low temperature,
spin degrees of freedom may be dropped.
Then the Hamiltonian in the plane reads
\begin{equation}
\label{H el}
{\cal H}_{\rm el} =
\sum_{ij, \nu\nu'} \tilde{t}^{\nu\nu'} c_{i\nu}^{\dagger} c_{j\nu'}
+
\tilde{U} \sum_{i} (n_{i1}-\frac{1}{2}) (n_{i2}-\frac{1}{2}),
\end{equation}
where $n_{i\nu} = c_{i\nu}^{\dagger}c_{i\nu}$ is a number operator.
The transfer matrices are renormalized independently of sites
because of the perfect polarization; and
the interaction $\tilde{U}$ is not a bare
interorbital Coulomb repulsion $U_{12}$
but $U_{12}-J_{12}$, where $J_{12}$ is the Hund's rule coupling
between orbitals $\nu=1$ and $2$.
Here hopping integrals between nearest-neighbor sites
in the 2D plane are explicitly given as
$\tilde{t}^{11} = -3\tilde{t}_{0}/4$,
$\tilde{t}^{22} = -\tilde{t}_{0}/4$,
$\tilde{t}^{12} = \tilde{t}^{21} = -(+) \sqrt{3}\tilde{t}_{0}/4$
in the $x(y)$ direction
%and $\tilde{t}^{11} = \tilde{t}^{12} = \tilde{t}^{21} = 0$,
%$\tilde{t}^{22} = -\tilde{t}_{0}$
%in the $z$ direction
for $d_{x^{2}-y^{2}} (\nu=1)$ and $d_{3z^{2}-r^{2}} (\nu=2)$ orbitals,
\cite{Anderson1959}
where $\tilde{t}_{0}$ is defined later.
This model can be viewed as a generalization of `ordinary' Hubbard models
where $t^{\nu\nu'} \propto \delta_{\nu\nu'}$.
\cite{Hubbard1963}
%Note that this model eq. (\ref{H el}) should be modified
%to take account of
%an antiferromagnetic stacking in the $z$ direction at half filling.
In the following, we consider the above 2D model
partly because of its numerical feasibility.
Although it ignores some of important effects
due to three-dimensionality,
it may contain at least nontrivial important physics of
the MI transition by hole doping
where the 2D anisotropy was suggested to be relevant.
\cite{Imada1998b}
Moreover,
%although an antiferromagnetic coupling in the $z$ direction
%is neglected,
this model is useful to discuss an orbital ordering
at half filling since it also has
2D structure as the spin ordering.
We also note that our model may have some relevance to
layered Mn compounds,
\cite{Moritomo1996}
which consists of stacking of single or double MnO$_{2}$ layers
with rather small crystal-field splitting of
$d_{x^{2}-y^{2}}$ and $d_{3z^{2}-r^{2}}$.

The rest terms in eq. (\ref{H tot})
denote the JT coupling in a classical treatment.
\cite{Millis1996b, Millis1996}
These should be appropriate
when lattice distortion has a long-ranged order.
In perovskite manganites,
a cooperative JT distortion which also has
2D anisotropy as spin and orbital ordering
appears at half filling.
\cite{Wollan1955, Goodenough1955, Elemans1971}
Because of the symmetry of $e_{g}$ wave functions,
we consider only two phonon modes.
\cite{VanVleck1939, Goodenough1963, breathing}
Then we obtain the forms as
%\begin{eqnarray}
\begin{equation}
\label{H el-ph}
%{\cal H}_{\rm el-ph} &=& \sum {\bf g}_{i} \cdot {\bf I}_{i} \\
%%
%{\cal H}_{\rm ph} &=& k \sum_{i} u_{i}^{2},
{\cal H}_{\rm el-ph} = \sum_{i} {\bf g}_{i} \cdot {\bf I}_{i},
\quad
{\cal H}_{\rm ph} = k \sum_{i} u_{i}^{2},
%\end{eqnarray}
\end{equation}
where ${\bf g}$ and ${\bf I}$ are two-component vectors defined as
%\begin{eqnarray}
$
{\bf g}_{i} = -2g u_{i} \left( \cos 2\theta_{i}, \sin 2\theta_{i} \right)
$
and
$
{\bf I}_{i} = \left( T_{i}^{z}, T_{i}^{x} \right)
$;
%\end{eqnarray}
$g$ and $k$ are the electron-phonon coupling and
the spring constant.
Here $u_{i}$ denotes displacement of oxygens surrounding the $i$-th Mn site
%and $\theta_{i}$ is defined below in eq. (\ref{gamma def}).
and $\theta_{i}$ determines the coupling angle.
$T_{i}^{\mu} (\mu=x,y,z)$ is called the pseudo-spin operator which denotes
two degrees of freedom of orbitals as
%\begin{equation}
$
%\label{T def}
T_{i}^{\mu} = \frac{1}{2} \sum_{\nu\nu'} \hat{\sigma}_{\nu\nu'}^{\mu}
c_{i\nu}^{\dagger} c_{i\nu'}
$,
%\end{equation}
where $\hat{\sigma}$ is the Pauli matrix.
%Using this definition, eq. (\ref{H el-ph}) is rewritten as
%\begin{equation}
%       \label{H el-ph:gamma}
%{\cal H}_{\rm el-ph} =
%-g \sum_{i} u_{i} \left( \gamma_{i1}^{\dagger} \gamma_{i1}
%                                    -\gamma_{i2}^{\dagger} \gamma_{i2} \right),
%\end{equation}
%where the operator $\gamma$ is a linear combination of the operator $c$ defined as
%\begin{equation}
%\label{gamma def}
%\left( \begin{array}{c}
%       \gamma_{i1} \\ \gamma_{i2}
%       \end{array} \right)
%=
%\left( \begin{array}{cc}
%          \cos \theta_{i} & \sin \theta_{i} \\
%          -\sin \theta_{i} & \cos \theta_{i}
%          \end{array} \right)
%\left( \begin{array}{c}
%          c_{i1} \\ c_{i2}
%          \end{array} \right).
%\end{equation}
%Therefore, this term eq. (\ref{H el-ph:gamma}) acts like
%a level splitting for two orbitals denoted by the new operators
%$\gamma_{\nu}$.

When the oxygen between the $i$ and $j$-th Mn sites moves
by $u_{i}$, the hopping integral between Mn sites changes
since originally it consists of a product of
overlap integrals between Mn and O sites.
Note that in both phonon modes,
oxygens shift only along the direction of Mn-O-Mn bonds.
A displacement of $u_{i}$ modifies the Mn-O hoppings
by a factor of $\left(1 \pm u_{i}\right)^{7/2}$,
\cite{Harrison1980}
therefore we define a unit of the Mn-Mn hopping integrals
in eq. (\ref{H el}) as
%\begin{equation}
%       \label{tildet0 def}
$
\tilde{t}_{0} = t_{0}
\left( 1+u_{i} \right)^{7/2} \left( 1-u_{i} \right)^{7/2}.
%\end{equation}
$

In the following, we discuss ground state properties
of the model eq. (\ref{H tot}) in two dimensions
by the projection QMC method.
\cite{White1989, Imada1989}
Computational details including the negative sign problem
will be described elsewhere.
\cite{MotomeUNPUBLISHED}
We take energy and length units as $t_{0}$ and
the lattice constant, respectively.
%Our Hamiltonian have many parameters;
%the Coulomb interaction $\tilde{U}$,
%the electron-phonon coupling $g$,
%the spring constant $k$,
%and parameters $u_{i}$ and $\theta_{i}$ which determine
%local structure of Jahn-Teller distortions.
From experiments and band calculations,
$t_{0}$ is considered to be around 0.5 eV
and $\tilde{U}$ is estimated at several eV.
\cite{Bocquet1992, Arima1993, Chainani1993, Satpathy1996, Solovyev1996a, Solovyev1996b}
In the present study, we change $\tilde{U}/t_{0}$ as a parameter
and discuss its realistic value based on our results.
The spring constant $k$ is roughly estimated by
the frequency of an oxygen bond stretching phonon at
order of $10$ or $100$ eV.
\cite{Millis1996b}
The electron-phonon coupling $g$ is not easy to determine experimentally,
however, it may be several eV.
\cite{Millis1996b}
We set here $k=100$ and $g=10$.
The parameters $u_{i}$ and $\theta_{i}$ are determined
by a mean-field scheme for simplicity;
we fix $\theta_{i}$ at experimental values as mentioned below
and determine a uniform solution $u_{i}=u$ to minimize total energy
obtained from the QMC calculations.
%As mentioned above,
%we note that couplings perpendicular to this plane work
%in a different way depending on hole concentrations;
%their effects will be discussed in elsewhere.
%\cite{MotomeUNPUBLISHED}

\begin{figure}
\hfil
\epsfxsize=7.5cm
\hfil
\epsfbox{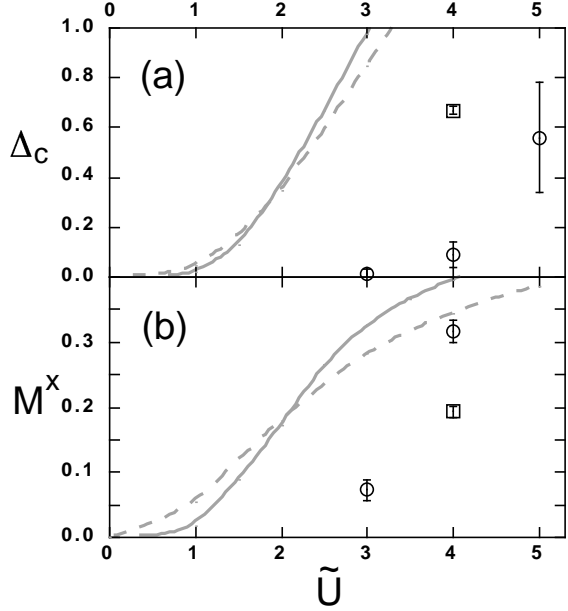}
%\epsfbox{fig1.EPSF}
\caption{
QMC results (circles) without JT coupling for
(a) the charge gap and
(b) the staggered orbital polarization.
The data are shown in the limit of $N_{\rm S}\rightarrow \infty$,
which are obtained by size extrapolations
from $N_{\rm S}=4\times 4$ to $10\times 10$.
Squares are QMC results for `ordinary' Hubbard models.
\cite{Assaad1996, White1989}
The gray (dotted) curve shows the mean-field results for our model
(for the `ordinary' Hubbard model).
}
\label{Fig: gap&Tx}
\end{figure}

First, we discuss results at half filling
($\sum_{i\nu} n_{i\nu} = N_{\rm S}$).
Fig. \ref{Fig: gap&Tx} shows the $\tilde{U}$ dependence of
charge gap and staggered moment of orbital polarization
in the absence of the JT coupling.
For the comparison,
QMC results for `ordinary' 2D Hubbard models are also plotted
\cite{Assaad1996, White1989}
with the mean-field results.
Here, the charge gap is calculated from shifts of chemical potential
upon doping.
\cite{Furukawa1992}
In our models, the charge gap opens
considerably slower than the case of `ordinary' Hubbard models,
which leads to a remarkable deviation from the mean-field results,
%especially in the small $\tilde{U}$ region,
presumably because of their characteristic nesting property.
\cite{MotomeUNPUBLISHED}
%Our model shows a perfect nesting property at half filling,
%therefore the system is considered to become an insulator
%at infinitesimal $\tilde{U}$ as `ordinary' Hubbard models.
%However, the nesting property has contrastive differences
%from that of `ordinary' Hubbard models.
%We speculate this may be a reason for the remarkable suppression
%of charge gap for small $\tilde{U}$.
In order to reproduce the realistic charge-gap amplitude
within the electronic model eq. (\ref{H el}),
we need a much stronger interaction.
Fig. \ref{Fig: gap&Tx} (b) shows that
the interaction $\tilde{U}$ induces
the staggered long-ranged orbital ordering of $T^{x}$ component, that is,
$(|d_{x^{2}-y^{2}}\rangle + |d_{3z^{2}-r^{2}}\rangle)
/(|d_{x^{2}-y^{2}}\rangle - |d_{3z^{2}-r^{2}}\rangle)$ type,
whose pattern is slightly different from the experimentally-observed
one, $|d_{3x^{2}-r^{2}}\rangle/|d_{3y^{2}-r^{2}}\rangle$.
This discrepancy has also been pointed out in previous studies
by the mean-field approximations.
\cite{Mizokawa1996, Shiina1997}
The moment grows rapidly as the values of $\tilde{U}$,
probably as a consequence of its uniaxial anisotropy.
Our results show that the Coulomb interaction $\tilde{U}$
by itself leads to a Mott insulating state with
a staggered orbital ordering, although
its ordering pattern does not show complete agreement with experiments.

\vspace*{-4mm}

\begin{figure}
\hfil
\epsfxsize=8cm
\hfil
\epsfbox{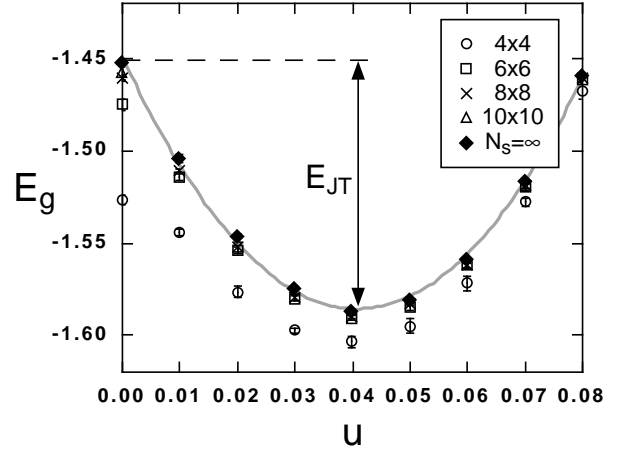}
%\epsfbox{fig2.EPSF}
\caption{
Ground state energy per site at half filling
as function of the oxygen distortion $u$.
We take $\tilde{U}=4$, $g=10$ and $k=100$.
The curve is a polynomial fit to the data at $N_{\rm S}=\infty$.
}
\label{Fig: JT GSE}
\end{figure}

If we take the JT coupling into account,
importance of electron correlation becomes clearer.
Fig. \ref{Fig: JT GSE} shows ground state energy per site as function of
oxygen displacement $u$.
Here we assume $\theta_{i} =  (-1)^{|{\bf r}_{i}^{x} + {\bf r}_{i}^{y}|} \pi/6$,
following the JT pattern observed in experiments.
The total energy is substantially lowered
and minimized at a finite $u$;
the stabilization energy
$E_{\rm JT} \simeq 0.13$ at $u \simeq 0.04$ for $\tilde{U}=4$.
At the same time, the orbital-ordering structure changes to
the experimentally-observed one,
namely $|d_{3x^{2}-r^{2}}\rangle/|d_{3y^{2}-r^{2}}\rangle$.
In addition, the charge gap amplitudes become considerably larger than
those without the JT coupling;
for instance,
%$\Delta_{\rm c} \simge 1$ for $\tilde{U} = 4$.
at $\tilde{U} = 4$, $\Delta_{\rm c} \simge 1$
($\sim 1.1$ for hole doping and 1.7 for electron doping).
These stabilization energy and charge gap grow
with increasing $\tilde{U}$ or $g/k$.
%Note that in contrast to this, $E_{\rm g}$ for
%a non-interacting system ($\tilde{U}$=0) has a minimum at $u=0$, that is,
%there is no instability to the Jahn-Teller distortion.
%\cite{bandcalc}
This indicates suppression of quantum fluctuations
as a cooperative and combined effect of $\tilde{U}$ and $g/k$.
Experimentally, $u$ is about $0.04$ and
$E_{\rm JT}$ is below or around 0.1 eV
in the light of the structural transition temperature
(for instance, $\sim 800$K for LaMnO$_{3}$).
\cite{Wollan1955, Goodenough1955, Elemans1971}
It is quite surprising that in spite of neglecting the three-dimensionality,
our present results show quantitative agreement with
these experimental results;
although more tuning of $\tilde{U}$ or $g/k$ 
(probably larger $\tilde{U}$) is required
to obtain more realistic value of the Mott gap.
We should stress that in the absence of $\tilde{U}$,
the JT distortion is far insufficient in reproducing
these results.
%Consequently, our results indicate not only 
%that the Coulomb interaction is essential
%to discuss the Jahn-Teller distortion and
%the orbital state at half filling;
%but also that our model with realistic parameters gives
%a consistent picture with experimental situations.
Synergetic effects of the strong correlation and the JT coupling
play a crucial role
to reproduce the realistic JT distortion,
orbital order configurations and the Mott gap amplitude.
\cite{bandcalc}

\vspace*{-5mm}

\begin{figure}
\hfil
\epsfxsize=8cm
\hfil
\epsfbox{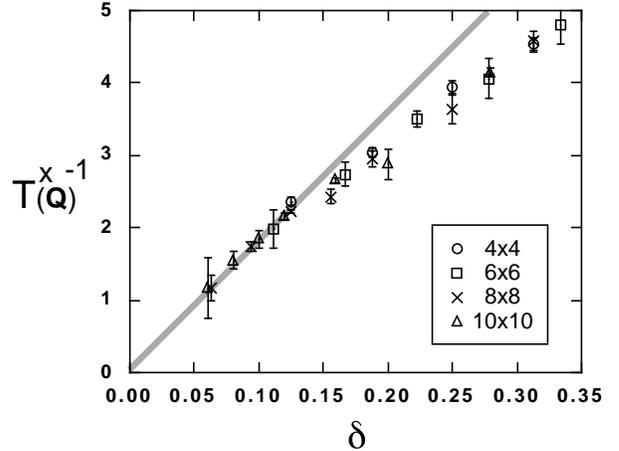}
%\epsfbox{fig3.EPSF}
\caption{
Inverse plot of peak values of the orbital correlation function
as function of doping concentration at $\tilde{U}=4$.
The gray line is the linear fit to the data at $0<\delta<0.15$.
}
\label{Fig: Tx vs. delta}
\end{figure}

\vspace*{-4mm}

Doped holes lead to an MI transition
from this Mott insulating state.
We focus effects of Coulomb interaction on this transition.
The ferromagnetic insulator presumably with the JT distortion
adjacent to the Mott insulator is beyond the scope of this paper
because the JT coupling was not studied in detail in doped cases.
Our data of doping concentration dependence of
chemical potential indicate that
there is no phase separation.
In Fig. \ref{Fig: Tx vs. delta}, we plot the inverse of
peak values of the orbital correlation function $T^{x}({\bf Q})$
where ${\bf Q} = (\pi,\pi)$
as function of hole concentration $\delta$.
Here we define
$
T^{\mu}\left({\bf k}\right) = N_{\rm S}^{-1} \sum_{ij}
T_{i}^{\mu} T_{j}^{\mu} \exp\left( {\rm i}{\bf k}{\bf r}_{ij}\right)
$
($\mu=x,y,z$).
Our results show that the staggered orbital correlations appear to be
insensitive to the system sizes for $\delta>0$.
Moreover, for small $\delta$, they are scaled as
$T^{x}({\bf Q})^{-1} \sim \delta$.
These indicate that
the staggered orbital polarization stabilized by $\tilde{U}$
is destroyed simultaneously with
the MI transition at $\delta=0$.
In `ordinary' Hubbard models, the similar scaling of spin correlations
and an anomalous scaling of compressibility were found.
\cite{Furukawa1992}
These suggest a novel universality class of the MI transition
where residual entropy of spin degrees of freedom
may induce strong incoherence of a metal near the critical point.
\cite{Imada1995}
In our model,
although we have not explicitly studied the compressibility anomaly yet,
the critical behavior of orbital correlations and
its implication of large entropy
may well be the origin of a strong incoherence observed in experiments.
\cite{Okimoto1995, Okimoto1997, Imada1998b}

In doped cases, our model exhibits
a uniform ordering of $d_{x^{2}-y^{2}}$ orbital,
\cite{MotomeUNPUBLISHED}
which has been suggested in recent experiments.
\cite{Kawano1998, Akimoto1998}
Our results show that a positive peak of
$T^{z}({\bf k}=0)$ grows as $\delta$ and
has a maximum around $\delta = 0.5$.
This ordering presumably comes from an optimization of 
the kinetic energy in the 2D planes,
since it is observed even in the non-interacting 2D model.
We note that this ordering is substantially enhanced
by $\tilde{U}$.

To summarize, we have investigated ground states of double-exchange models
with on-site Coulomb interaction,
twofold degeneracy of conduction bands and JT coupling
using the QMC method
under the condition of perfect-polarization of spins in 2D planes.
Experimental results of undoped manganites, especially the JT distortion,
are quantitatively reproduced
with realistic parameters.
Importance of the synergy between the strong electron correlation
and the JT coupling is clarified.
Upon doping, our models show a continuous metal-insulator transition
with a critical enhancement of staggered orbital correlations.
We have speculated its relation to strong incoherence of charge dynamics
in ferromagnetic metals observed in experiments.
In addition, our models exhibit a uniform ordering of
$d_{x^{2}-y^{2}}$ orbital in doped cases,
which agrees with experimental indications.

The authors thank Hiroki Nakano for fruitful discussions
and suggestive comments.
This work is supported
by `Research for the Future Program from Japan Society
for Promotion of Science (JSPS-RFTF 97P01103).
Y. M. is supported by Research Fellowships of
Japan Society for the Promotion of Science for Young Scientists.
A part of the computations in this work was performed
using the facilities of the Supercomputer Center,
Institute for Solid State Physics, University of Tokyo.


\begin{thebibliography}{99}
\bibitem{breathing}
Upon doping, another mode in which oxygens surrounding a hole
are breathing may be important.
\end{thebibliography}
\end{document}